\documentclass[10pt,preprintnumbers,aps,amssymb,nofootinbib,amsmath]{revtex4}
\usepackage{epsfig,epsf}
\usepackage{graphicx}
\usepackage{epstopdf}
\newcommand{\beq}{\begin{equation}}
\newcommand{\eeq}{\end{equation}}
\def\eq#1{{(\ref{#1})}}

\def\re#1{{Ref.~\cite{#1}}}

\newcommand{\as}{\alpha_s}
\newcommand{\un}{\underline}

\newcommand{\im}{\mathrm{Im}}
\begin{document} 

%\preprint{Draft \#1}

\title{Heavy quark production in Color Glass Condensate}

\author{ Kirill Tuchin$\,^{a,b}$}
\affiliation{
$^a\,$Department of Physics and Astronomy, Iowa State University, Ames, IA 50011 \\
$^b\,$RIKEN BNL Research Center,
Upton, NY 11973-5000}

\date{\today}

\begin{abstract} 
We discuss heavy quark production in High Parton Density QCD in quasi-classical approximation and including low-$x$ quantum evolution. We also consider an alternative approach based on the effect of pair production in external fields.
\end{abstract}

\begin{center}

\maketitle 
\emph{
Based on talk presented at the International Conference on Strangeness in Quark Matter, University of California, Los Angeles, Mach 30, 2006.
}\end{center}

%%%%%%%%%%%%%%%%%%%%%%%%%%%%%%%%%%%%%%%%

\section{Quasi-classical approximation }\label{intro}

Production of a quark $q$ and antiqiuark $\bar q$ pair at high energies is characterized by two time scales: production time $\tau_P$ and  interaction time $\tau_\mathrm{int}$.  In pA collisions in the center-of-mass frame of the $q\bar q$ pair the production time is $\tau_P\simeq 1/(2m)$, where 
$m$ is a quark's mass. In the nucleus rest frame, this time is Lorentz time-dilated by $E_g/(2m)$ where $E_g$ is energy of a gluon in a proton from which the $q\bar q$ originates. In  terms of the Bjorken variable $x_2=(m_T/\sqrt{s}) e^{-y}$ the production time is $\tau_P\simeq1/(2 M x_2)$, where  
$M$ is a nucleon mass. At RHIC this corresponds to the production time $\tau_P\simeq 15e^y$~fm. On the other hand, the interaction time is $\tau_\mathrm{int}\simeq R_A\simeq 7$~fm.  Clearly, in the limit $y\gg 1$ one can
consider the pA interaction as an instantaneous process. This implies in particular, that in the configuration space the transverse coordinates of quarks and gluons do not change over time of interaction in which case color dipoles diagonalize the interaction matrix. 

The time structure of heavy quark production discussed in the previous paragraph can be taken into account in the light-cone perturbation theory. It is convenient to use the nucleus rest frame. In \re{Tuchin:2004rb} a model was considered in which 
$q\bar q$ pair appears as a result of the following chain of fluctuations in a fast proton: $q_v\rightarrow g\rightarrow q\bar q$, where $q_v$ is a valence quark and $g$ is a gluon. Thus, the heavy quark production amplitude consists of  contributions from three different time orderings: (i) valence quark interacts with the nucleus and then emits a gluon, (ii) gluon interacts with the nucleus and then emits a $q\bar q$ pair, (iii)  $q\bar q$ interacts with the nucleus. Correspondingly, there are six different time orderings in the cross section \cite{Tuchin:2004rb}. The light-cone wave function for each time ordering was calculated in \re{Kovchegov:2006qn}. In addition, one also has to calculate the propagator of a parton system  for each time ordering.  The propagator sums up multiple scattering of a dipole in a nucleus. In the quasi-classical approximation this amounts to resuming higher order corrections in $\as$ enhanced by powers of large nuclear length $\as A^{1/3}\sim 1$. 
 Explicit expressions for all these propagators can be found in \re{Tuchin:2004rb,Kovchegov:2006qn}.

 The final result for the  double-inclusive
quark--anti-quark production cross section in $pA$ collisions in the
quasi-classical approximation is \cite{Kovchegov:2006qn}
\beq\nonumber
\frac{d \, \sigma}{d^2k_1 \, d^2 k_2 \, dy \, d \alpha \, d^2 b} \,=\, 
\frac{1}{4 \, (2 \, \pi)^6} \, \int d^2 x_1 \, d^2 x_2 \, d^2 y_1 \, d^2 y_2 \, 
e^{- i \, \un k_1 \cdot (\un x_1 - \un y_1) - i \, \un k_2 \cdot (\un x_2 - \un y_2)} 
\eeq
\beq\label{dcl}
\times \, \sum_{i,j=1}^3\, \Phi_{ij} (\un x_1, \un x_2; \un y_1, \un y_2; \alpha) \
\Xi_{ij} (\un x_1, \un x_2; \un y_1, \un y_2; \alpha),
\eeq
where $\un k_1$, $\un k_2$ are the quark and anti-quark transverse momenta, $y$ is quark's rapidity, $\alpha$ is a fraction of gluon's energy carried by a quark, $\un b$ is impact parameter.  
$\Phi_{ij}$ and $\Xi_{ij}$ ($i,j=1,2,3$) are light-cone wave functions and propagators for different time orderings in configuration space. Integrations in \eq{dcl} go over all transverse coordinates in a proton wave function.  

The complicated result of \re{Kovchegov:2006qn} can be simplified for the aligned jet configuration:  when quark or anti-quark takes away most of the gluon's energy ($\alpha\ll 1$) \cite{Tuchin:2004rb}.  In that case the gluon production time becomes much larger than that of $q\bar q$ which allows factorization of the light-cone wave functions into those of gluon $\Phi_{g\rightarrow q\bar q}$ and quark--anti-quark $\Phi_{q_v\rightarrow q_v\bar g}$. 
 Then the differential cross section for single inclusive quark production in pA collisions reads   \cite{KopTar,Tuchin:2005ky,Blaizot:2004wv}
\begin{eqnarray}\label{main}
&&\frac{d\sigma}{d^2k \,dy\,d^2b}=
\frac{1}{2(2\pi)^4} \int d\alpha\,
\int d^2x\, d^2 y\, d^2x'\, e^{-i\un k\cdot (\un x-\un y)}\,
\Phi_\mathrm{g\rightarrow q\bar q}(\un x,\un x',\un y, \alpha)\,
\Phi_\mathrm{q_v\rightarrow q_v\bar g}(\un x,\un x',\un y, \alpha)\,
\nonumber\\
&\times&  \bigg(
e^{-\frac{1}{4}\, \frac{C_F}{N_c}\, (\un x-\un y)^2\, Q_s^2}\,-\,
e^{-\frac{1}{4}\, \frac{C_F}{N_c}\, (\un x-\un x') ^2\, Q_s^2}-
e^{-\frac{1}{4}\, \frac{C_F}{N_c}\, (\un y-\un x')^2\, Q_s^2}+1\bigg)\,,
\end{eqnarray}
where I assumed for simplicity that the dominant contribution comes from 
interaction of the $q\bar q$ pair with the target, while rescatterings of 
the gluon and the valence quark are neglected. However in general, they 
must be taken into account as well. The ligh-cone wave functions in \eq{main} are given by  
\beq\label{wf1}
\Phi_\mathrm{q_v\rightarrow
q_v\bar g}(\un x,\un x',\un y, \alpha)
\,=\,\frac{\as\, C_F}{\pi^2}\,
\frac{(\alpha\un x+(1-\alpha)\un x') \cdot(\alpha\un y+(1-\alpha)\un x')}
{(\alpha\un x+(1-\alpha)\un x')^2\,(\alpha\un y+(1-\alpha)\un x')^2}\,,
\eeq
$$
\Phi_{g\rightarrow q\bar q}(\un z,\un x, \un x', \alpha)\,=\,
\frac{\as}{\pi}\, m^2\,\bigg(\, \frac{(\un x-\un x')\cdot
(\un y -\un x')}{|\un x-\un x'|\, |\un y -\un x'|}   
K_1(|\un x-\un x'|\,m)\,K_1(|\un y -\un x'|\,m)
$$
\beq\label{wf2}
\times[\,\alpha^2\,+\,(1-\alpha)^2\,]
+\,K_0(|\un x-\un x'|\,m)K_0(|\un y -\un x'|\,m) \,\bigg)\,,
\eeq
where $Q_s$ is the saturation scale, $m$ is a quark mass. 

The gluon saturation scale $Q_s$ is given by
\beq
Q_s^2 \, = \, \frac{2 \, \pi^2 \, \as }{C_F} \, \rho \, T(\un b)\, xG_N(x,1/\un x^2)\,,
\eeq
where $T(\un b)$ is a nucleus profile function and $\rho$ is the nuclear density. 
The saturation scale measures the strength of the nuclear color field at high energies  $F\sim Q_s^2/g$ (see reviews \cite{Iancu:2003xm,Jalilian-Marian:2005jf}). 

%%%%%%%%%
\section{Including quantum evolution}

When the interaction energy becomes such that $\as y\sim 1$ it becomes necessary to include the effects of quantum evolution at small $x$ \cite{BK} into the $q\bar q$ pair production cross section \cite{Blaizot:2004wv,Kovchegov:2006qn}.  This is done by first generalizing the quasi-classical model to include possible gluon emission from valence anti-quark. Thus, we get a formula for  the case of $q\bar q$ production in dipole--nucleus scattering. 
Strictly speaking, our results would then only be applicable to
particle production in deep inelastic scattering. However, our results
below may still serve as a good approximation for gluon production in
$pA$ collisions \cite{Kovchegov:2001sc}.  
If the transverse
coordinates of the quark and anti-quark in the incoming dipole are
denoted by $\un z_0$ and $\un z_1$ correspondingly with $\un z_{01} =
\un z_0 - \un z_1$, we write
$$
\frac{d \, \sigma}{d^2k_1 \, d^2 k_2 \, dy \, d \alpha \, d^2 b} (\un z_{01}) \,=\, 
\frac{1}{4 \, (2 \, \pi)^6} \, \int d^2 x_1 \, d^2 x_2 \, d^2 y_1 \, d^2 y_2 \, 
e^{- i \, \un k_1 \cdot (\un x_1 - \un y_1) - i \, \un k_2 \cdot (\un x_2 - \un y_2)} 
$$
\beq\label{dcl_dip}
\times \, \sum_{i,j=1}^3\, \sum_{k,l=0}^1 \, (-1)^{k+l} \, \Phi_{ij} (\un x_1 - \un z_k, 
\un x_2 - \un z_k; \un y_1 - \un z_l, \un y_2 - \un z_l; \alpha) \
\Xi_{ij} (\un x_1, \un x_2, \un z_k; \un y_1, \un y_2, \un z_l; \alpha),
\eeq
with generalized propagators $\Xi_{ij}$  \cite{Kovchegov:2006qn}.

The inclusion of quantum corrections due to leading logarithmic
(resumming powers of $\as \, y$) approximation in the large-$N_c$
limit is done along the lines of \cite{Kovchegov:2001sc} (see also \cite{Jalilian-Marian:2005jf}
 for a review) using Mueller's dipole 
model formalism \cite{dipole}. Since the integration over rapidity
interval separating the quark and the anti-quark in the pair does not
generate a factor of the total rapidity interval $Y$ of the collision
(i.e., does not give a leading logarithm of energy), the prescription
for inclusion of quantum evolution is identical to the single gluon
production case. We first define the quantity $n_1 ({\un z}_{0}, \un
z_1; {\un z}_{0'}, \un z_{1'}; Y-y)$, which has the meaning of the
number of dipoles with transverse coordinates ${\un z}_{0'}, \un
z_{1'}$ at rapidity $y$ generated by the evolution from the original
dipole ${\un z}_{0}, \un z_1$ having rapidity $Y$. It obeys the dipole
equivalent of the BFKL evolution equation \cite{dipole,BFKL}
$$
\frac{\partial n_1 ({\un z}_{0}, \un
z_1; {\un z}_{0'}, \un z_{1'}; y)}{\partial y} \, = \, 
\frac{\as \, N_c}{2 \, \pi^2} \, 
\int d^2 z_2 \, \frac{z_{01}^2}{z_{20}^2 \, z_{21}^2} \, 
\bigg[ n_1 ({\un z}_{0}, \un
z_2; {\un z}_{0'}, \un z_{1'}; y) + 
n_1 ( {\un z}_{2}, \un
z_1; {\un z}_{0'}, \un z_{1'}; y) 
$$
\beq\label{eqn}
- n_1 ({\un z}_{0}, \un z_1; {\un z}_{0'}, \un z_{1'}; y) \bigg] 
\eeq
with the initial condition 
\beq
n_1 ({\un z}_{0}, \un z_1; {\un z}_{0'}, \un z_{1'}; y=0) \, = \,
\delta ( \un z_0 - \un z_{0'} ) \, \delta ( \un z_1 - \un z_{1'} ).
\eeq
If the target nucleus has rapidity $0$, the incoming dipole has
rapidity $Y$, and the produced quarks have rapidity $y$, the inclusion
of small-$x$ evolution in the rapidity interval $Y-y$ is accomplished
by replacing the cross section from \eq{dcl_dip} by \cite{Kovchegov:2001sc}
\beq\label{sub1}
\frac{d \, \sigma}{d^2k_1 \, d^2 k_2 \, dy \, d \alpha \, d^2 b} (\un z_{01}) 
\rightarrow \int d^2 z_{0'} \, d^2 z_{1'} \, n_1 ({\un z}_{0}, \un z_1; 
{\un z}_{0'}, \un z_{1'}; Y-y) \, 
\frac{d \, \sigma}{d^2k_1 \, d^2 k_2 \, dy \, d \alpha \, d^2 b} (\un z_{0'1'}).
\eeq

Inclusion of evolution in the interval between $0$ and $y$ is
accomplished by replacing the Mueller-Glauber rescattering exponents
according to the following rule \cite{Kovchegov:2001sc}
\beq\label{sub2}
e^{- \frac{1}{4} \, (\un x_0 - \un x_1)^2 \, Q_s^2 \, 
\ln (1/|\un x_0 - \un x_1| \, \Lambda)} \, \rightarrow \,
1 - N (\un x_0, \un x_1, Y)
\eeq
where $N (\un x_0, \un x_1, Y)$ is the forward amplitude for a quark
dipole $\un x_0, \un x_1$ scattering on a target with rapidity
interval $Y$ between the dipole and the target. (We refer the interested reader to the \re{Kovchegov:2006qn} for explicit expressions of the resulting propagators.) 
The forward scatetring amplitude  obeys the following
evolution equation \cite{BK}
\beq
\frac{\partial N ({\un x}_{0}, {\un x}_1, Y)}{\partial Y} \, = \, 
\frac{\as \, N_c}{2 \, \pi^2} \, 
\int d^2 x_2 \, \frac{x_{01}^2}{x_{20}^2 \, x_{21}^2} \, 
\left[ N ({\un x}_{0}, {\un x}_2, Y) + 
N ({\un x}_{2}, {\un x}_{1}, Y) - N ({\un
x}_{0}, {\un x}_1, Y) \right. 
\eeq
\beq\label{eqN}
- \left. N ({\un x}_{0}, {\un x}_{2}, Y) \, N ({\un x}_{2}, {\un x}_{1}, Y)
\right]
\eeq
with the initial condition
\beq
 N ({\un x}_{0}, {\un x}_1, Y=0) \, = \, 1 - e^{- \frac{1}{4} \, (\un
 x_0 - \un x_1)^2 \, Q_s^2 \, \ln (1/|\un x_0 - \un x_1| \, \Lambda)}.
\eeq

The final result for  the double inclusive $q\bar q$ production cross section
including small-$x$ evolution effects reads
$$
\frac{d \, \sigma}{d^2 k_1 \, d^2 k_2 \, dy \, d \alpha \, d^2 b} (\un z_{01}) 
\, = \, \frac{1}{4 \, (2 \, \pi)^6} \, 
\int d^2 z_{0'} \, d^2 z_{1'} \, n_1 ({\un z}_{0}, \un z_1; 
{\un z}_{0'}, \un z_{1'}; Y-y) \,
$$
$$
\times \,  d^2 x_1 \, d^2 x_2 \, d^2 y_1 \, d^2 y_2 \, 
e^{- i \, \un k_1 \cdot (\un x_1 - \un y_1) - i \, \un k_2 \cdot (\un x_2 - \un y_2)} 
$$
\beq\label{dcl_ev}
\times \, \sum_{i,j=1}^3\, \sum_{k,l=0}^1 \, (-1)^{k+l} \, \Phi_{ij} (\un x_1 - \un z_k, 
\un x_2 - \un z_k; \un y_1 - \un z_l, \un y_2 - \un z_l; \alpha) \
\Xi_{ij} (\un x_1, \un x_2, \un z_k; \un y_1, \un y_2, \un z_l; \alpha, y).
\eeq
Integrating over one of
the quarks' transverse momenta we obtain the single inclusive quark
production cross section
$$
\frac{d \, \sigma}{d^2 k \, \, dy \, d^2 b} (\un z_{01}) 
\, = \, \frac{1}{2 \, (2 \, \pi)^4} \, 
\int d^2 z_{0'} \, d^2 z_{1'} \, n_1 ({\un z}_{0}, \un z_1; 
{\un z}_{0'}, \un z_{1'}; Y-y) \, d^2 x_1 \, d^2 x_2 \, d^2 y_1 \,  
e^{- i \, \un k \cdot (\un x_1 - \un y_1) }
$$
\beq\label{single_ev}
\times \, \int_0^1 d\alpha \sum_{i,j=1}^3\, \sum_{k,l=0}^1 \, (-1)^{k+l} \, 
\Phi_{ij} (\un x_1 - \un z_k, 
\un x_2 - \un z_k; \un y_1 - \un z_l, \un x_2 - \un z_l; \alpha) \
\Xi_{ij} (\un x_1, \un x_2, \un z_k; \un y_1, \un x_2, \un z_l; \alpha, y).
\eeq

These equations generalize the results Refs.~\cite{KopTar,Tuchin:2004rb,Gelis:2003vh,Blaizot:2004wv}.

%%%%%%
\section{Schwinger mechanism}

Alternative approach to particle production in high energy hadron and nuclei collisions was advocated in Refs.~\cite{Kharzeev:2005iz,Kharzeev:2006zm}. It was argued there that the particle production from the Color Glass Condensate can be considered as a vacuum instability in strong background longitudinal color field. It has been known for a long time that strong longitudinal color fields may be important for particle production in  hadronic reactions \cite{Casher:1978wy} and in particular, in heavy-ion collisions \cite{Biro:1984cf,Kajantie:1985jh,Bialas:1986mt,Kerman:1985tj,Gatoff:1987uf,Kluger:1991ib}. However, the structure of these fields remained unknown due to their non-perturbative nature characterized by large coupling $\as\sim 1$. The Color Glass Condensate eliminates  this problem  by introducing the hard scale $Q_s\gg \Lambda_\mathrm{QCD}$ so that the classical color fields can be in principle calculated as a series in small coupling and high density of charges
$\as(Q_s^2)\,\rho\sim 1$, where $\rho \propto A^{1/3}$. Existence and dominance of \emph{longitudinal} fields in color glass was realized in \cite{Kharzeev:2005iz,Kharzeev:2006zm} and later in \cite{Fries:2005yc}. 

Moreover, according to the parton model, particles in a parton cascade are strongly ordered in energy. This implies that the production times of a partons in a cascade are also strongly ordered. Therefore, it makes sense to consider motion of a slow parton in an almost static field of faster partons. One can also argue that the transverse sizes of partons are strongly ordered as well, at least in the leading logarithmic approximation, which makes the longitudinal chromoelectric field $E$ the dominant in the target rest frame \cite{Kharzeev:2006zm}.

Consider for simplicity motion of a scalar charged particle $\phi$ in the background field $A_\mu$
\beq \label{EQM}
  (\partial_\mu \,-\,i g A_\mu)^2\, \phi\,=\,0\,.
\eeq 
The field is given by  $E_z=-\partial_- A_+=\mathrm{const}$.
 Although $A_+$ is a function of only $x_-$ Eq.~\eq{EQM} cannot be solved by separation of variables since the initial condition depends on both $x_+$ and $x_-$. Indeed, a fast hadron \emph {interacts} with a target located at, say, 
$z=t=0$.  Instantaneously, the parton cascade looses its coherence. Therefore, 
we have to solve equation of motion \eq{EQM} with the constraint that the potential $A_+$ vanishes at  $z= 0$, i.e.\ explicitly depends on both lightcone coordinates $x_+$ and $x_-$.  
Thus, we are looking for the solution in the form 
$\phi(x) =e^{-i S-ip_\bot \cdot x_\bot}$. 
Working in the WKB approximation $|\partial_+ S \partial_-S|\gg |\partial_+\partial_- S|$ we reduce \eq{EQM} to 
\beq\label{sem}
-2\partial_+ S (\partial_- S -g A_+(x_-)) + p_\bot^2 + 2g\,\sigma E_z  =0\,, \quad x_+\ge x_-\,,
\eeq
where $\partial_+=\frac{\partial}{\partial x_-}$ and $\partial_-=\frac{\partial}{\partial x_+}$.
Eq.~\eq{EQM} is a Hamilton-Jacobi equation of motion of a charged particle in the background field $A_+=A_-(x_+)\theta(x_+-x_-)$.  We can write and solve the Hamilton equations of motion for the canonical momenta 
$p_-=-\partial_- S$, $ p_+=-\partial_+ S$ as a functions of the light-cone coordinates. The result is \cite{Kharzeev:2006zm}
\begin{eqnarray}
p_+&=& -gA_+(x_-)\,+\, gA_+(x_+)\,+\,p_+^0\,,\label{brb1}\\
x_-&=& \frac{p_\bot^2}{2}\int\frac{dx_+}{(p_+^0+gA_+(x_+))^2} \,,\label{brb2}\\
S&=& -\int gA_+(x_-)dx_- \,+\, \int dx_+\frac{p_\bot^2}{p_+^0+gA_+(x_+)}\label{brb3}\,.
\end{eqnarray}
Eq.~\eq{brb2} coincides with  the equation of motion of a classical test particle of mass $p_\bot$ in the external field $A_+(x_+)$.  In other words,  the test particle effectively moves under the action of the longitudinal electric field $E_z=-A_+'(x_+)$. 

Eq.~\eq{brb3} gives the action of the test particle along the trajectory \eq{brb2}. Its imaginary part arises from the pole in the integrand of the second term in the right-hand-side of \eq{brb3}.
Integration around the pole in the plain of complex $x_+$ yields the imaginary part. It can be calculated replacing the denominator in the first integral   in \eq{brb3} by $\im (p_+^0+gA_+)^{-1}= \pm (i \pi/2) \delta(p_+^0+gA_+)$ according to  the Landau-Cutkosky cutting rule. Additional factor of $1/2 $ arises due to the condition $x_+\ge x_-$. Define 
\beq\label{somedef}
\tau\,=\, x_+\omega\,,\quad A_+(\tau)\,=\,-\frac{E_0}{\omega}f(\tau)\,,\quad 
\gamma\,=\, \frac{p_+^0\omega}{gE_0}\,.
\eeq
where $\omega$ is a typical frequency of the external field and $E(\tau=0)=E_0$. With this definitions we obtain
\beq\label{import}
\im S\,=\, \im\int \frac{p_\bot^2}{gE_0}\frac{d\tau}{\gamma -f(\tau)}\,=\,
\frac{p_\bot^2}{2gE_0}\frac{\pi}{f'(f^{-1}(\gamma))}\,.
\eeq
The imaginary part of the action \eq{import} corresponds to the pair production. 

The physical meaning of the \emph{adiabaticity parameter} $\gamma$ introduced in \eq{somedef} is clear: $\gamma=0$ for the static field, 
while $\gamma\gg 1$ for rapidly oscillating one. Since $gE_0\simeq k_{i,+}^2$ and $\omega=k_{i,-}$, with subscript $i$ denoting a parton making up the background field,
 we have the following estimate
\beq\label{nv2}
\gamma\simeq\,\frac{p_+}{k_{i,+}}\,.
\eeq
This estimate implies  
that  due to the strong ordering of the light-cone momenta in the partonic cascade at $\tau=0$,  the emission of the gluons is determined by small values of $\gamma$ or,
 in other words, by constant  electric fields, in which $A_+ (x_+)\,= E_0 \,x_+$.
Using this formula in \eq{import} yields the result  for pair production probability \cite{Schwinger}
\begin{equation}\label{sch}
w\simeq \exp\left\{-\frac{\pi (m^2+p_\bot^2)}{gE}\right\}\,.
\end{equation} 
We would emphasize that \eq{sch} is not analytic at small coupling which is a manifestation of its non-perturbative nature. At small transverse momenta it has a finite limit even at $m=0$. 
Therefore, the proposed method of calculating the pair production at high energies allows access to the kinematical region of small $p_\bot$ where the perturbation theory breaks down.  

As later times the background field decays due to the effect of  screening by produced pairs which results in modification of the spectrum \eq{sch} \cite{Kharzeev:2005iz,Kharzeev:2006zm}. 

%%%%%%%%%%%%%%%%%%%%%%%%%
\section{Phenomenological applications}

Equations \eq{dcl_ev}, \eq{single_ev} and \eq{sch}  have important
phenomenological applications for studying the dense partonic system
in p(d)A and eA collisions. Observation of hadron suppression in the
nuclear modification factor measured in dA collisions at forward
rapidities at Relativistic Heavy Ion Collider (RHIC) \cite{dAdata}
signals the onset of the nonlinear evolution of the scattering
amplitude for light hadrons \cite{KLM,Kharzeev:2003wz,KW}. Due to a
large mass, the impact of nonlinear evolution effects on the heavy
quark production is shifted to higher energy and/or rapidity. It was
estimated in \cite{KhT} using the $k_T$-factorization approach (see also \cite{Fujii:2006ab}) that
one can expect a significant deviation of the open charm production
cross section from the perturbative behavior already at
pseudo-rapidity $\eta\simeq 2$ at RHIC. Due to the heavy quark
production threshold one expects that the total multiplicity of
open charm scales as $N_\mathrm{coll}$ at lower energy and/or
rapidity whereas at higher energies and/or rapidities the scaling
law should coincide with that for lighter hadrons \cite{KhT}, i.\
e.\ open charm multiplicity should scale as $N_\mathrm{part}$
\cite{KL} due to high parton density effects.  Therefore, to be able
to compare predictions of CGC with the data reported by RHIC
experiments and to make predictions for the possible upcoming $pA$ run
at the Large Hadron Collider (LHC), it is important to perform a
calculation of an open charm production within the more general
approach developed in this paper. Our final results
\eq{dcl_ev}, \eq{single_ev} and \eq{sch} allow one to describe open charm
transverse momentum spectra at different rapidities and center-of-mass
energies, allowing for a complete description of RHIC and LHC
data. Since the saturation scale $Q_s$ is expected to be even higher
at LHC than it was at RHIC, the CGC effects on heavy quark production
at LHC should be even more significant.

%%%%%%%%%%%%%%%%%%%%%%%%%%%%%%%%%%%%%%%%

\vskip0.3cm
{\bf Acknowledgments}
%\vskip0.3cm
The author is grateful  to  Dima Kharzeev, Yuri Kovchegov and Genya Levin for fruitful collaborations.  The author would like to thank
RIKEN, BNL and the U.S. Department of Energy (Contract
No.~DE-AC02-98CH10886) for providing the facilities essential for the
completion of this work.

%%%%%%%%%%%%%%%%%%%%%%%%%%%%%%%%%%%%%%%%%%%%%%%%%%%%%%%%%%%%%%%%%%%%%%%%%%%%

\end{document}